\documentclass[structabstract]{aa}
\usepackage[OT1]{fontenc}
\usepackage[latin9]{inputenc}
\usepackage{booktabs}
\usepackage{textcomp}

\usepackage{amsmath}
\usepackage{txfonts}
\usepackage{graphicx}
\usepackage{amssymb}
\begin{document}

\title{Sunyaev-Zel'dovich galaxy clusters number counts :  consequences of  cluster scaling laws evolution}
\titlerunning{SZ number counts}

\author{Pierre Delsart\inst{1},
 	Domingos Barbosa\inst{2},
	\and Alain Blanchard\inst{1}
}

\authorrunning{P. Delsart, D. Barbosa \& Alain Blanchard}

%\offprints{Pierre Delsart}

\institute{Laboratoire Astrophysique de Toulouse-Tarbes, UMR 5572, Universit\'e de Toulouse, 14 rue Edouard Belin, 31400 Toulouse, France\\
\email{pierre.delsart@ast.obs-mip.fr,alain.blanchard@ast.obs-mip.fr}
\and
Instituto de Telecomunica\c c\~oes, Universidade de Aveiro, Campus Universitario de Aveiro, 3810-183 Aveiro, Portugal\\ 
\email{dbarbosa@av.it.pt}
}

\abstract {}
{Galaxy cluster surveys based on the Sunyaev-Zeldovich effect (SZE) mapping are expected from ongoing experiments. Such surveys are anticipated to provide a significant amount of information relevant to cosmology from the number counts redshift distribution. We carry out an estimation of predicted SZE counts and their redshift distribution taking into account the current cosmological constraints and the X-ray cluster temperature distribution functions. Comparison between local and distant cluster temperature distribution functions provides evidence for an evolution in the abundance of X-ray clusters that is not consistent with the use of standard scaling relations of cluster properties in the framework of the current concordance model. The hypothesis of some evolution of the scaling law driven by non-gravitational processes is a natural solution to this problem.}
{We perform a MCMC statistical study using COSMOMC, combining current CMB observations from WMAP, the  SNIa  Hubble diagram, the galaxy power spectrum data from SDSS and X-ray clusters temperature distributions  to predict SZE cluster number counts.} 
{Models reproducing well the X-ray cluster temperature distribution function evolution lead to a significantly lower SZE clusters number counts with a distinctive redshift distribution. Ongoing microwave SZE surveys will therefore shed new light on intracluster gas physics and greatly help to identify the role of possible non-gravitational physics in the history of the hot gas component of x-ray clusters.}
{}


\maketitle

\section{Introduction}

Knowledge on the galaxy cluster population has greatly progressed, thanks both to targeted observations and to systematic surveys carried by space X-ray facilities. The hot intra-cluster gas is also known to interact with the incoming photons of the Cosmic Microwave Background (CMB), leaving a specific frequency imprint  known as the thermal Sunyaev-Zeldovich effect (\cite{1972CoASP...4..173S}). This change of the sky brightness of the CMB can be written as a function of frequency and the Compton parameter, $y$, proportional to the integrated gas pressure along the line-of-sight of the cluster. For an individual cluster, the integrated Compton parameter $Y$ is the value of $y$ integrated over the solid angle subtended by the clusterwhich angular size happens to be nearly independent of redshift. This turns the SZE effect in a very effective probe of clusters at cosmological distances. Furthermore, SZE selection is very attractive in principle as the signal depends on an integral of the intracluster gas pressure, independently of its spatial distribution.

The relevance of these SZE cluster surveys for cosmological application has been outlined in the past (\cite{1994ApJ...423...12B}, \cite{1994ApJ...426....1M}, \cite{1996A&A...314...13B}), motivating the onset of key science projects for some of the most exciting experiments in the microwave regime. In particular, the Planck Surveyor ESA mission and the South Pole Telescope (\cite{2004SPIE.5498...11R}) (SPT hereafter) are crucial experiences to explore the SZE number counts and their cosmological significance. 
%Planck Surveyor is the third generation microwave and millimetre waves space mission, aiming the production of the  first all-sky SZE catalogue at moderate resolution, discovering several thousands of previously unknown clusters. SPT is carring a blind SZE survey over 4000$\rm deg2$ and is expected to discover many previously undetected thousands of clusters with a mass selection criteria quite uniform with redshift (\cite{2005A&A...429..417M}).
 With redshifts obtained from the optical and infrared follow-up, such samples 
%catalogues 
have the potential to complement other cosmological probes and constrain 
  the matter density parameter, $\Omega_M$,  the amplitude of matter 
fluctuations measured by the parameter $\sigma_8$, the dark energy content
 of the Universe, and its equation of state $w$ (\cite{2007PhRvD..76l3013L}). 
Besides these important cosmological inferences, the SZE cluster cartography 
may also reveal the cluster formation processes out to high redshifts, an 
aspect which is further explored in the present letter.

\section{Cluster modelling}

The modelling of the clusters population and its evolution needs two key ingredients. The first one is an expression of the mass function and its dependence on cosmology. This was first theoretically attempted by Press and Schechter in 1974 (\cite{1974ApJ...187..425P}). Since that time, numerical simulations have shown that the mass function follows a simple and nearly universal scaling relation with a dependence actually close to the initial  Press and Schechter proposition, with deviations much below the precision level needed for present day applications (\cite{2008ApJ...688..709T}). The second key ingredient is the relation between cluster mass and observable quantities. From scaling arguments (\cite{1991ApJ...383..104K}), we can relate the temperature measured from X-ray to the mass:
\begin{equation}\label{eq:TM}
T_X=A_{TM}(hM)^{2/3}\left( \Omega_M \frac{\Delta(z,\Omega_M)}{178} \right)^{1/3}(1+z)
\end{equation}
in which $\Delta(z,\Omega_M)$ represents the density contrast (relative to the background density)  by which clusters are defined, $h$ is the Hubble parameter and $A_{TM}$ describes the normalization of the $T_X-M$ relation. Hydrodynamical simulations of cluster formation have provided mass-temperature relations that actually follow the above relationship with dispersion of the order of 20\%, while the normalization constant $A_{TM}$ is uncertain depending on the gas physics (\cite{2007MNRAS.377..317K}). The magnitude of the SZE is controlled by $Y$, the integrated Compton parameter depending on the gas mass and the average gas temperature:
\begin{equation}\label{eq:YM}
Y=KM_gT_gD^{-2}_a
\end{equation}
where $D_a$ is the angular distance, $M_g$ the total gas mass of the cluster, $T_g$ its (mass-)average temperature and $K$ is a normalization depending only on physical constants. This relation, being independent of the actual spatial distribution of the hot gas, makes the SZE signal an appealing proxy of the total mass. If the gas was isothermal in clusters, one would have $T_g$ = $T_X$. However, it is known that clusters are not isothermal with the temperature declining in outer parts (\cite{2005ApJ...628..655V}) following a scaling law. We can therefore still assume that the gas temperature follows some scaling laws but with a different normalization $T_g = \xi T_X$. Using the observed gas fraction in clusters (\cite{2005A&A...437...31S}), a dark matter  mass profile according to numerical simulation of cold dark matter (\cite{1997ApJ...490..493N}) and the observed temperature profile (\cite{2005ApJ...628..655V}) one can estimate $\xi \sim 0.6$, a value which is quite uncertain, but in the following this number is left as a quantity to be determined from observations. With the above mass-temperature relation, the $Y-M$ relation can be evaluated numerically:

\begin{equation}\label{eq:YMfull}
Y=1.816.10^{-4}\xi A_{TM}f_B M^{5/3}h^{8/3} \left( \Omega_M \frac{\Delta(z,\Omega_M)}{178} \right)^{1/3}(1+z) D^{-2}
\end{equation}

with $f_B$ as the baryonic gas fraction, $D$ the dimensionless part of the angular distance and $Y$ expressed in $\rm arcmin^2$. In this work, we take $f_B$ as the universal baryonic fraction corrected for some possible depletion $\Upsilon$: $f_B=\Upsilon \frac{\Omega_b}{\Omega_M}$ . For a typical Coma-like cluster with about $10^{15}$ solar masses, observed at $z=1$, and using the standard scaling laws, the typical Compton distortion is $Y$ is of the order of $2.3\times 10^{-4} \rm arcmin^2$, corresponding at 143GHz to a flux decrement with respect to the mean CMB flux of -21mJy, or a temperature fluctuation of -650$\mu$K, well above both Planck and SPT sensitivity thresholds.
Assuming the above relations, we can now compute the SZE source counts as well as their redshifts distribution, as expected in a Friedmann-Lema\^{i}tre cosmological model: theoretical number counts predictions of cosmic objects can be obtained from the mass function describing the formation and evolution of objects over redshift. In this way, cluster counts $dN(z)$ for clusters with intrinsic $Y$ greater than a sensitivity threshold $Y_0$ are described  as:
\begin{equation}\label{eq:dndz}
\frac{dN(z)}{dz} = \int_{M_0(z)}^{+\infty}\frac{dV}{dz}\frac{dn}{dM}dM
\end{equation}
where the mass threshold $M_0(z)$ is obtained from equation (\ref{eq:YMfull}) $dV$ is the volume element, accounting for the geometry and expansion rate of the Universe, and $\frac{dn}{dM}$ is the cluster mass function, depending on the matter and energy densities of the universe and on the initial power spectrum of mass fluctuations. On the other hand, the very same SZE cluster population is a source of non-gaussian secondary CMB anisotropies (\cite{1988MNRAS.233..637C}, \cite{1999ApJ...526L...1K}, \cite{2005ApJ...626...12B}, \cite{2010MNRAS.404.1197T}), connecting the present large scale structure to the primordial CMB spectrum. This offers a way to pin down the very nature of the astrophysical evolution with redshift of the cluster $Y-M$ relation. 

\section{The X-ray picture}

\begin{figure}

\includegraphics[width=0.4\paperwidth,height=0.25\paperheight]{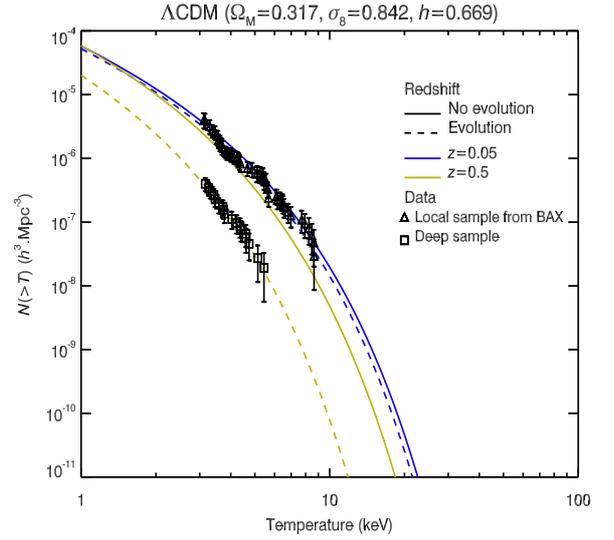}

\caption{Cluster temperature distribution function in the local and distant universe. Triangles are our estimation of the  temperature distribution function. 
The stars represent the estimated temperature distribution function at high redshift from the 400$\rm deg2$ sample restricted to the redshift range [0.4-0.6]. The continuous line corresponds to the predicted temperature distribution adjusted to fit CMB, SNIa, LSS and the local temperature distribution function. The predictions for this model at redshift 0.5 (the lower continuous line) are far in excess of the abundances estimated from observations. For visual comparison, the dashed lines are the corresponding predictions with an evolving mass-temperature relation, eq (6), with $\alpha = -1$ (not fitted).}

\label{xray}
\end{figure}
However, since cluster masses are hardly measurable directly, it is necessary to use a relation between  mass with some other observables, including its dispersion, to derive constraints from the observed temperature distribution function of X-ray clusters (\cite{1992A&A...262L..21O}). The constraints read as a thin region in the parameter space ($A_{TM}$, $\sigma_8$) (\cite{2003MNRAS.342..163P}), a known degeneracy between $A_{TM}$ and $\sigma_8$ clearly identified and quantified. In the present analysis we use the standard $\Lambda$-Cold Dark Matter picture ($\Lambda$CDM) and Bayesian Monte Carlo Markov Chain (MCMC) estimations. We use the COSMOMC parameters estimation package with constraints provided by CMB data from WMAP 7 years (\cite{2010arXiv1001.4744J}), Supernovae SN Ia data from the SDSS compilation sample (\cite{2009ApJS..185...32K}), matter power spectrum $P(k)$ estimation from SDSS LRG DR7 (\cite{2010MNRAS.404...60R}) and our estimation of  cluster temperature distribution at different redshifts. 
For this, we use the local temperature distribution function of X-ray clusters from a local sample of  clusters with fluxes above $2\times10^{-11}\rm erg.s^{-1}.cm^{-2}$ in the [0.1-2.4] keV band and with galactic latitude $|b|> 20$ degrees obtained from the cluster X-ray data base BAX. This results in a sample comprising $48$ clusters. The theoretical temperature distribution function was derived from the mass function with the above defined mass temperature relation, using the Sheth, Mo and Tormen mass function  (\cite{2001MNRAS.323....1S}). We can then constraint the six cosmological parameters of this vanilla model ($\Omega_M$, $\Omega_B$, $\tau$, $h$, $n$, $\sigma_8$) and the normalization $A_{TM}$ of the mass-temperature relation. As an illustration, we provide the theoretical temperature distribution function computed for the median model (the normalization being $A_{TM} = 7.31$ keV, Figure \ref{xray}), compared to the observed local abundance of X-ray clusters. Clearly, this model reproduces well the local abundance of clusters. A previous claim that the predicted abundance of X-ray clusters in a standard concordance cosmological model exceeds the observed abundance at large redshift led us to compute   the  high redshift ($z\sim 0.5$) temperature distribution function of clusters in the best concordance model and compare it to the one inferred from the high-redshift subsample of the 400$\rm deg2$ survey (\cite{2009ApJ...692.1033V}). We find the standard scaling relation vastly overpredicts the derived X-ray cluster distribution function at high redshift. If we allow for some redshift evolution to compare the predicted abundance of X-ray clusters a mass temperature relation, we may rewrite the $T_X-M$ relation as :

\begin{equation}\label{eq:TMe}
T_X=A_{TM}(hM)^{2/3}\left( \Omega_M \frac{\Delta(z,\Omega_M)}{178} \right)^{1/3}(1+z)^{(1+\alpha)}
\end{equation}

Such non standard scaling law has been advocated (\cite{2003A&A...412L..37V}) with $\alpha \sim -1$ in order to match the abundance of high redshift X-ray clusters in a concordance cosmology in flux selected surveys. The  predicted temperature distribution function at high redshift with non-standard evolution of the mass-temperature evolution (using a power-law index of $\alpha= -1$ without further adjustment) is clearly preferred over the non evolving one. Potential explanations for such evolution are likely to lie in non-gravitational heating of the intra-cluster gas, possibly tracing sources of cosmic stellar formation or/and additional complex physical processes within clusters, an area that is the subject of intense investigations (\cite{2005RvMP...77..207V}, \cite{2005Natur.433...45M}, \cite{2007MNRAS.376.1547C}, \cite{2009MNRAS.396..849D}).

\section{The anticipated Sunyaev-Zel'dovich number counts}

\begin{figure*}
%\includegraphics[width=0.8\paperwidth,height=0.4\paperheight]{delsartfig2.ps}
%\begin{center}
\includegraphics[scale=.925]{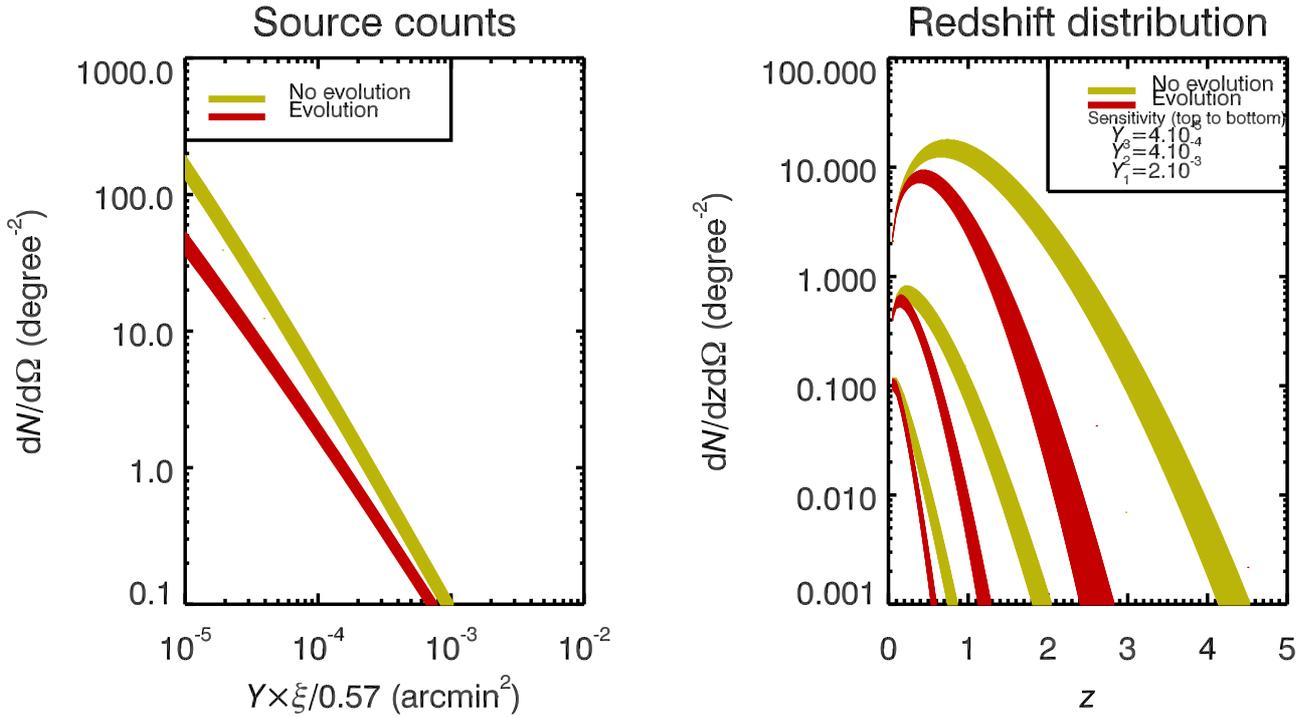}
%\end{center}
\caption{Predicted galaxy clusters number counts detected from the Sunyaev-Zeldovich effect, after joint analysis of CMB, SNIa, LSS data and estimated temperature distribution function. Models describing self consistently the low and high redshift temperature distribution functions, needing non-standard evolution of $T_X-M$ relation, lead to lower abundance of clusters, showing a deficit in the cluster number increasing with redshift (right).}

\label{sz}
\end{figure*}

An early attempt to estimate SZE number counts using the mass function and scaling laws as well as an estimation of their contribution to CMB fluctuations on small scales has been performed by Bartlet and Silk (\cite{1994ApJ...423...12B}). The use of SZE counts as a cosmological probe (\cite{1996A&A...314...13B}) was proposed around the early framework of Planck Surveyor Phase A studies, and soon followed by detailed investigations (\cite{1997A&A...325....9A}, \cite{2000ApJ...544..629H}, \cite{2001MNRAS.325..835K}, \cite{2002MNRAS.331...71B}, \cite{2002MNRAS.331..556D}) further exploiting the potential of SZE surveys counts as a cosmic probe. Such estimations were usually performed under the assumption of the standard scaling law of the mass temperature relation. Since the SZE number counts are by definition the number of galaxy clusters brighter than a certain flux threshold $S_{\nu}$, this reduces to the number of galaxy clusters more massive than a mass threshold $M_0$. depending on the redshift $z$, from the $Y-M$ relation. The consequence of the dispersion is dealt by assuming that in practice it can be approximated by a shift in the mass-temperature normalization (\cite{2000A&A...362..809B}).
In order to work out a consistent modelling of the clusters population and its evolution in the same picture, another MCMC analysis was performed including the redshift evolution power-law index $\alpha$ as an additional free parameter and adding the high redshift temperature distribution functions as a complementary observational constraint. For each of these models, we can compute the corresponding SZE source counts as well as their redshift distribution of the clusters, allowing a comparison of the importance of evolution effect to the statistical uncertainty inherent to the set of observations under consideration.
Here, we evaluate expected high-sensitivity flux limits for both Planck and SPT as fiducial numbers. An important source of complexity, and possibly uncertainty in this topic is the accuracy of our knowledge of the experiment selection functions (\cite{2005A&A...429..417M,2007A&A...465...57J}). 
We focus on the theoretical curves, leaving to the future works the full understanding of the selection functions used to find clusters which will need a comprehensive modeling of the observational and detection processes, including the understanding of the known complexities of the missions algorithms to identify clusters and measure its fluxes. Here as fiducial criteria, we adopted three different sensitivity thresholds: $Y = 2.10^{-3}$ (Case 1), corresponding to a secure lower limit for the sensitivity of Planck, $Y = 4.10^{-4}$ (Case 2) for SPT or for an optimistic flux limit for Planck, $Y = 4.10^{-5}$ (Case 3) which could be achieved if the selection function is well under control for experiments like SPT (\cite{2004SPIE.5498...11R}) or AMIBA (\cite{2009ApJ...694.1610H}). 
From the above discussion, at the brightest fluxes, source counts should be relatively free from resolution and/or selection effects, and allow some detection of the signature of evolution. Our final predicted counts and the associated uncertainties are shown in Figure 2: we give the predicted counts falling in the central 68\% interval using cosmological models fitted to the local temperature distribution function assuming standard scaling law for the mass-temperature relation; on the same figure, we provide expected counts for self consistent modelling of the cluster population, allowing evolution of the mass-temperature relation and including the high redshift temperature distribution function data as an additional constraint. It is clear that the integrated source counts show significant differences between the two cases. The possible lack of accurate knowledge of the $Y-M$ normalization and of the selection function is likely to limit the interpretation of SZE integrated counts. However, the redshift distribution is definitively different, with a clear drop of the high redshift tail after $z>1$ when the X-ray motivated evolution is taken into account. Because the SZE effect is sensitive to the actual average temperature of the gas, the observations in SZE will allow to  clarify whether the kind of evolution needed for reproducing X-ray number counts affects the overall cluster gas or only the central region from which the X-ray emission comes. This is rather encouraging as the high redshift clusters are likely to be point like in a moderately low resolution survey as will be provided by Planck, and thus alleviating the complex and possibly unsecured modelling of their spatial distribution for the determination of the selection function.

\section*{Conclusion}

%SZE number counts are potentially sensitive to both the cosmological framework and to the physics of the hot gas content of clusters. 
Current cosmological data and existing surveys of X-ray clusters provide tight constraints and thereby accurate predicted SZE number counts.
% , anticipating that forthcoming SZE surveys will provide key information for the understanding of the intracluster gas heating history.
 The X-ray data probe an inconsistency between the theoretically expected and the observed high redshift temperature distribution function under the current vanilla $\Lambda$CDM cosmology if the  standard scaling of the mass-temperature relation is used. The hypothesis of some evolution of the scaling law driven by non-gravitational processes is a natural solution to this problem. Our MCMC statistical study shows that evolution is required and that such evolution lead to a significant lowering of  the anticipated  SZE cluster number counts. Indeed, this may explain part of the statistical discrepancies suggested by South Pole (\cite{2009arXiv0912.4317L}) on the presence of a SZE signal anisotropy contribution to CMB fluctuations lower than expected in $\Lambda$CDM. This may also result from the geometrical structure of clusters (\cite{2010MNRAS.404.1197T}). While the X-ray data allowed the highlight of a redshift evolution in the galaxy cluster population, the SZE number counts have the potential to provide deeper insight on the actual nature of this evolution, by providing information allowing to test whether 
%on the overall cluster. 
%For the typical high sensitivities probed by Planck and SPT, our modelling shows that expected numbers significantly differ when evolution is included, this difference being enlightened at high redshift. If the ultimate Planck Surveyor and South Pole Telescope mission legacy catalogues reveal less high-$z$ clusters than expected from modeling using normal scaling relations, this will be a serious indication that 
gas clusters did undergo a significant amount of non-gravitational heating affecting their global energy budget. This would call for further dedicated investigations in future analysis on cluster abundance and formation with potential consequences on the design of future experiments. 

\begin{acknowledgements}
We acknowledge discussions with N.~Aghanim, M.~Douspis, J.~Silk and A.~da Silva. DB is supported by FCT Ci\^encia 2007 program funded through QREN and COMPETE. DB and PD were sponsored by FCT project PTDC/CTE-AST/65925/2006. We acknowledge the use of Cosmological MonteCarlo (COSMOMC) software package for statistical analysis. This research has made use of the X-Rays Clusters Database (BAX) which is operated by the Laboratoire d'Astrophysique de Tarbes-Toulouse (LATT). We acknowledge the use of the Legacy Archive for Microwave Background Data Analysis (LAMBDA). We acknowledge Universit\'e Paul Sabatier for the use of the cluster Hyperion.
\end{acknowledgements}
%============
% BIBLIOGRAPHY
%============

\end{document}